\documentclass[pre,onecolumn]{revtex4}
\usepackage{graphicx}
\usepackage{amsopn}
\usepackage{amsfonts}
\usepackage{latexsym}
\usepackage{amsmath,bm,amssymb}
\usepackage{color}
\usepackage[cp1251]{inputenc}

\begin{document}

\title{Bird's-eye view to the realm of $N_{TB}$ liquid crystals}

\author{E.I.Kats}

\affiliation{Landau Institute for Theoretical Physics, RAS, \\ 
142432, Chernogolovka, Moscow region, Russia}

\begin{abstract}
In this work we start with an introductory review of the relatively recently discovered type of liquid crystals, twist-bend nematics $N_{TB}$.
Then we describe a more detailed account of recent developments in the field.
Namely: (i) Landau Theory of the easy axis $N_{TB}$ nematics; (ii) Light scattering in the $N_{TB}$ liquid crystals;
(iii) Rheological behavior of the ordered $N_{TB}$ samples; (iv) Shift of the $N$ - $N_{TB}$ phase transition point
(where $N$ stands for the conventional nematic liquid crystal)
under actions of external fields.  
It is found that advocated in this work simple phenomenological approach 
successfully explains available experimental data.
The paper integrates the input from publications of other researchers, and presents 
a personal view of the author formed partially by his own results and discussions with his collaborators.

\end{abstract}

\date{}

\maketitle

\section{Introduction.}
\label{intr}
Discovered more than 100 years ago liquid crystals (aka - meso-phases, i.e., intermediate between isotropic fluids and
crystalline solids) are nevertheless much younger than other equilibrium states of matter known from ancient times \cite{SD04}.
Thus it is not surprising that there is no yet an exhaustive list of all possible types of liquid crystals.
Of course the reason is more deep that only a relatively young age of liquid crystalline science. The matter is that
there is only a single macroscopic symmetry group $O(3) \times T(3)$ of all isotropic liquids (where $O(3)$ is three-dimensional
(3D) rotation group, and $T(3)$ is 3D translation group). Similarly all possible crystalline solids may
be classified by their symmetry according to one of the complete set of 230 space symmetry groups (also known as Fedorov's groups).
On the contrary there is infinite number of possible symmetry groups for liquid crystals (e.g., any order rotational axis are not forbidden,
whereas translational symmetry of crystalline solids allows only 2, 3, 4, and 6 rotational axis).

Irrespective of said above until relatively recent time (the first decade of XXI - th century) it was believed that only
a few liquid crystalline structures are realized in Nature. Namely, possessing full 3D translational symmetry, 
nematics \cite{GP93} - \cite{BL11} (with $D_{\infty h}$ symmetry group 
for the uniaxial nematics and $D_{2 h}$ symmetry group for the biaxial nematics), chiral nematics or cholesterics 
with $C_{\infty }$ symmetry. 
Here and in what follows we use so-called Schoenflies notations for the macroscopic symmetry classes: $D_{\infty h}$
stands for the uniaxial, infinite order rotation axis, supplemented by the orthogonal to the axis mirror plane;
$D_{2 h}$ stands for the 2-d order rotation axis supplemented by the orthogonal to the axis mirror plane, and $D$ is replaced
by $C$ for the chiral (without space inversion symmetry) systems. The uniaxial nematics ordering traditionally is characterized by the unit
vector, director ${\bf n}$ and for non-polar liquid crystals the states ${\bf n}$ and $-{\bf n}$ are equivalent. 

Besides there exist also a number of smectic and discotic liquid crystals,
where mentioned above orientation symmetry is combined with 2D or 1D translational symmetry. Therefore smectic and discotic liquid 
crystals are more abundant than fully fluid-like nematics. The most well studied and known are smectics A ($Sm A$) liquid crystals with
equidistant fluid-like planes. Hence, the $Sm A$ symmetry is $D_{\infty h} \times T(2) \times L(1)$, where $L(1)$ stands for 1D
crystalline lattice with a simple (single period) periodicity. $Sm A^*$ is the chiral counterpart of $Sm A$. Its symmetry is 
$C_\infty \times T(2) \times L(1)$, but except its chirality $Sm A^*$ is indistinguishable from the $Sm A$ liquid crystals.

Smectics with anisotropic layers are known as smectics C ($Sm C$). There exist two, say microscopic, causes for the layer
anisotropy. First one is related to the molecular (director) tilt with respect to the normal ${\bf l}$ to the layer. In this case,
the layer anisotropy is determined by the vector ${\bf c}$ (it is a genuine vector, ${\bf c}$ and $-{\bf c}$ is not equivalent,
although it is termed traditionally by c-director)
\begin{eqnarray}
\label{r1}
{\bf c} = {\bf n} - {\bf l}({\bf n} {\bf l})
.
\end{eqnarray}
Note the symmetry under transformation ${\bf n} \to -{\bf n}$ and simultaneously ${\bf l} \to -{\bf l}$, and the fact
that $|{\bf c}| \neq 1$ (it determines the molecular tilt angle).
The second cause for the layer anisotropy is related to biaxiality of the orientation order. In this case we have a quadrupole layer
anisotropy, determined by the 2D second order symmetric and traceless tensor $Q_{ik}^{(b)}$
\begin{eqnarray}
\label{r2}
Q_{ik}^{(b)} = s_1(m_im_k - t_it_k) + s_2 (m_i t_k + m_k t_i)
,
\end{eqnarray}
where $s_1$ and $s_2$ are two scalar quantities which determine the degree of the biaxiality within the smectic layers, and three mutually
orthogonal unit vectors
${\bf n}$, ${\bf m}$ and ${\bf t}$ define the coordinate frame.

Due to their already existing and potential applications, special attention has been attracted to chiral liquid crystals.
Star superscript is traditionally used to distinguish achiral and chiral phases, e.g., $N^*$, $Sm A^*$, $Sm C^*$, and so on.
The matter is that a number of new effects can be expected (and some of those are observed) in chiral liquid crystals.
For example, an electroclinic effect in $Sm A^*$, or improper ferroelectricity in $Sm C^*$ phase, to name a few.
These known effects \cite{GP93}-\cite{OP06} are related in smectics to broken
symmetry within the smectic layers. Another option, namely, modulation of smectic layer periodicity, leads
to a multiple types of so-called smectic subphases \cite{DZ12} -  \cite{DK13}, \cite{HW15}, \cite{FV21}.

It is evident from aforesaid that physics of liquid crystals is far from being exhausted. A lot of obtained results
and observations is still awaiting for their classifications, systematization and pedagogical descriptions. 
However the aim of this paper is not to review of all known (or possible) types of liquid crystals. The aim is much more modest.
In what follows
we will not touch the classical (''old'') liquid crystals (nematics, cholesterics, smectics and discotics), unless it is needed to compare with
the new liquid crystals. Neither smectic subphases will not be considered here (all the more that relatively recent and details review
papers are available \cite{DZ12} - \cite{FV21}, and we refer readers who are interested in subphases to these review articles.

In this paper we will review structural and physical properties of only one new (relatively recently discovered) type of liquid crystals, so-called twist-bend nematics $N_{TB}$.
The paper integrates the input from publications of other researchers, and presents 
a personal view of the author formed partially by his own results and discussions with his collaborators. 
Correspondingly to the topic, in the next section \ref{prel} we describe the general features of the $N_{TB}$ nematics.
Then in section \ref{landau} we present Landau Theory of the easy axis $N_{TB}$ nematics.
Section \ref{optics} is devoted to a description of light scattering in the $N_{TB}$ liquid crystals.
In section \ref{rheology} we discuss rheological behavior of the ordered $N_{TB}$ samples. 
Then in section \ref{electric} we analyze how external d.c. electric field shifts $N$ - $N_{TB}$ phase transition point.
We end the paper with some conclusions and brief discussions of still open questions and perspectives.

\section{Preliminaries on the existence of modulated nematics}
\label{prel}

We start our paper with a description of relatively recent discovery \cite{HS09,PN10,CD11,BK13,MD14} of so-called
twist-bend nematics, ($N_{TB}$). $N_{TB}$ structure exhibits helical (thus chiral) orientation ordering despite being formed from achiral molecules. The $N_{TB}$ structure is essentially different from the known more than a century chiral cholesteric structure. The latter
one possess  simple (orthogonal) helical structures with pitches in a few $\mu m$ range. 
By its nature the orientation order parameter in the cholesteric phase is long-wavelength order parameter, locally identical with the nematic 
order parameter (it is characterized by slightly biaxial quadrupole second order tensor with the unit headless director  $\bf n$ being
along the long axis of this tensor).

In standard thermotropic cholesterics (or chiral nematics) there is no temperature driven phase transition between chiral and
achiral (nematic) structures. The both structures are locally identical (up to a small on the order of $a q_0$ factor, where $a \simeq 10^{-7}\,
cm$ is molecular scale, and $q_0 \simeq 10^{-4}\, cm^{-1}$ is the wave vector of cholesteric spiral modulation). Over the same mall parameter
$a q_0 \ll 1$ th order parameter in cholesterics are long wavelength second order tensor $Q_{ik}$ (similarly to
the nematic liquid crystals). This cholesteric like modulated structure occurs at any (sufficiently small) $q_0$ value, characterizing
the system chirality. $N_{TB}$ orientation order is basically different, and the parameter $a q_0 \simeq 1$. The short wavelength
$N_{TB}$ orientation order is basically different, and the parameter $a q_0 \simeq 1$. The short wavelength
modulation in the $N_{TB}$ structure suggests that the order parameter should be also short wavelength one (unlike long wavelength
director ${\bf n}$ or second order tensor $Q_{ik}$ introduced above). The difference between the both types of the order parameters
are not semantic one. The long wavelength order parameters by its definition, in becomes soft (or what is the same, easily
excitable) in the limit when the wave vector $q$ of the excitation tends to zero $q \to 0$. In more technical terms it means
for example that the system free energy in the vicinity of the phase transition can be expanded over the order parameter itself,
and also over the gradients of the order parameter (corresponding small parameter is $a q_0$). It is not 
the case for short wavelength order parameter (when $a q_0 \simeq 1$). In the vicinity of the $N\, -\, N_{TB}$ phase transition,
the $N_{TB}$ order parameter becomes soft at the finite value $q_0$ of the wave vector. The free expansion over the gradients of the short wavelength order parameter must take into account this fact ($a q_0 \simeq 1$). Instead of the expansion over the small $q$ (small gradients
$(\nabla Q_{ik})^2 \ll 1$) 
of the long wavelength order parameter, it should be an expansion over small deviations from the principle short wavelength modulation 
vector $q_0 \simeq a^{-1}$. For example in the isotropic case, the corresponding gradient term is proportional to $(\nabla ^2 + q_0^2)^2$.

With all said above in mind, we assume that a system (liquid crystal) is deeply in the nematic phase, where director $\bf n$
provides soft degrees of freedom related to nematic ordering. In the vicinity of the phase transition into the
modulated $N_{TB}$ phase. Since the modulation in the $N_{TB}$ structure is short wavelength, the corresponding
order parameter is also the short wavelength one.
Therefore the free energy expansion should include the following terms \cite{KL14}, \cite{KA17}
\begin{itemize}
\item
Conventional, long wavelength Frank energy
\begin{eqnarray}
 {\cal F}_\mathrm{Fr}=\int dV\
 \left\{ \frac{K_1}{2} (\nabla\cdot\bm n)^2
 +\frac{K_2}{2}\left[\nabla_\perp n_i
 \nabla_\perp n_i
 \right. \right.\nonumber \\ \left. \left.
 -(\nabla\cdot\bm n)^2\right]
 +\frac{K_3}{2} (\partial_z \bm n)^2 \right\}.
 \label{Frank}
 \end{eqnarray}
\item
Short-scale components of the order parameter should be treated in terms of different from (\ref{Frank}) elastic energy. 
We introduce a short-scale component $\bm\varphi$ of the order parameter. 
The components have to be orthogonal to the long wavelength director, $\bm n\cdot \bm \varphi=0$. The vector $\bm\varphi$ has two independent components. The quantity $\bm\varphi$ plays a role of the order parameter for the phase transition $N$--$N_{TB}$. 
The corresponding Landau functional in terms of $\bm\varphi$ may not contain odd over this order parameter terms
(since $\varphi$ is a vector). Therefore in the mean field approximation the $N$--$N_{TB}$ transition is a continuous (second order) phase transition, and the Frank energy (\ref{Frank}) should be supplemented by its short wavelength counterpart.
\item
Defined above the short wavelength vector order parameter ${\bm \varphi }$ is orthogonal to the long wavelength
director ${\bf n}$. However, the wave vector of the modulation ${\bf q}$ could be oriented arbitrary in space.
We consider three possible types of the wave vector orientation. Namely, along the director (easy-axis, one dimensional
modulation), perpendicular to the director (easy-plane, two dimensional modulation), and mixed state with
three dimensional tilted with respect to the director modulation wave vector.

\end{itemize}

Non-gradient terms of the Landau free energy expansion for short wavelength order parameter have
the same universal as for usual long wavelength order parameter
\begin{eqnarray}
F_0 =
 \int dV \left\{\frac{a}{2} \bm\varphi^2
 +\frac{\lambda}{24} {\bm \varphi}^4
 \right\}, 
 \label{lr1}
 \end{eqnarray}
since ${\bm \varphi }$ is a vector, odd order terms are forbidden.

On the contrary, the gradient terms for the short wavelength order parameter are different
from those for the long wavelength order parameter. Moreover, the terms depend on the orientation
of the modulation wave vector (the vector order parameter ${\bm \varphi}$ itself is always perpendicular to the
director). According to this criterion, it is convenient to classify all $N_{TB}$
structures into three groups:
\begin{itemize}
\item
For the easy-axis structure, the modulation wave vector oriented along the director (cf. with anisotropic
magnetic structures, where the classification is based on the orientation of the magnetization (order parameter) vector).
Then the main term describing the softening of the
order parameter in the vicinity of two points ${\bf q} = \pm \kappa _{||} {\hat e}_z$ (where ${\hat e}_z$ is the unit vector
along $Z$ axis chosen along the non-perturbed director ${\bf n}$) reads as
\begin{eqnarray} 
 \int dV \left\{\frac{b_{||}}{8 \kappa _{||}^2} \left[
 \left(n_i n_k \partial_i \partial_k + \kappa _{||}^2\right) \bm\varphi \right]^2
  \right\} .
 \label{lr2}
 \end{eqnarray}
However because for the short wavelength order parameter ${\bm \varphi }$ its
gradients are not small, the contributions (\ref{lr1}) should be supplemented by the following
2-d order terms
\begin{eqnarray} 
 \int dV \left\{\frac{b_1 }{2} (\nabla \bm\varphi)^2
  +\frac{b_2}{2}\delta^\perp_{ij}
 \partial_i \bm \varphi \partial_j \bm \varphi \right \}
 , 
 \label{lr3}
 \end{eqnarray}
with where $\delta^\perp_{ij}=\delta_{ij}-n_i n_j$. The terms (\ref{lr3}) are the same order as the quadratic term $\propto a{\bm \varphi }^2$.
Similarly the 4-th order term
\begin{eqnarray} 
 \int dV \left\{ 
\frac{\lambda_1}{16 \kappa _{||}^2}
 \left(\epsilon_{ijk} \varphi_i \partial_j \varphi_k\right)^2
 \right\}, \quad
 \label{lr4}
 \end{eqnarray}
is generally of the same order as the term $(\lambda /24){\bm \varphi }^4$.
Combining everything together we end up with the following minimal model for the $N$ to the easy-axis $N_{TB}$
phase transition
\begin{eqnarray} 
 \int dV \left\{\frac{a}{2} \bm\varphi^2
 +\frac{b_{||}}{8 \kappa _{||}^2} \left[
 \left(n_i n_k \partial_i \partial_k +\kappa _{||}^2\right) \bm\varphi \right]^2
 +\frac{b_1}{2} (\nabla \bm\varphi)^2
 \right. \nonumber \\ \left.
 +\frac{b_2}{2}\delta^\perp_{ij}
 \partial_i \bm \varphi \partial_j \bm \varphi
 +\frac{\lambda}{24} \varphi^4
 -\frac{\lambda_1}{16 \kappa _{||}^2}
 \left(\epsilon_{ijk} \varphi_i \partial_j \varphi_k\right)^2
 \right\}, \quad
 \label{bana1}
 \end{eqnarray}
 As usual, $a\propto T-T_c$, where $T_c$ is the mean field transition temperature. 
The quantities $b$ are analogs of the Frank moduli for the order parameter $\bm\varphi$. The free energies (\ref{Frank}) and
(\ref{bana1}) is the minimal Landau model for the $N$--easy-axis $N_{TB}$ phase transition.
\item
For the transition $N$ into the easy plane $N_{TB}$ phase (as above the state is defined
by the orientation of the modulation wave vector, not by the orientation of the vector order
parameter) the softening of the order parameter ${\bm \varphi }$ occurs near
the circle $|{\bf q}| = \kappa _\perp $, and the term (\ref{lr2}) should be replaced by the following one (with the self-evident
change of the notations)
\begin{eqnarray} 
 \int dV \left\{\frac{b_{\perp}}{8 \kappa _{\perp}^2} \left[
 \left(\partial_\perp ^2 +\kappa _{\perp}^2\right) \bm\varphi \right]^2
  \right\} ,
 \label{lr5}
 \end{eqnarray}
where subscript $\perp $ is defined in the plane orthogonal to the director ${\bf n}$ (e.g., $\nabla _\perp ^2 \equiv \nabla ^2
-({\bf n} \nabla )^2$.
Other terms entering the minimal model $(\ref{bana1})$ free energy, have the same form (again up to the
change of notations) as for the easy-axis $N_{TB}$. Namely, (\ref{lr3}), (\ref{lr4}) read for the easy plane $N_{TB}$ as
\begin{eqnarray} 
 \int dV \left\{\frac{b_3 }{2} (\nabla _\perp \bm\varphi)^2
  -\frac{\lambda _2}{16 \kappa _\perp ^2}
 [(\varphi_i \nabla _i) \varphi _k]^2 \right \}
 .
 \label{lr6}
 \end{eqnarray}
\item
When the modulation wave vector is tilted with respect to director $\bf n$ (in what follows the case will be termed three dimensional
or tilted $N_{TB}$ phase), the softening of the vector order parameter ${\bm \varphi }$ occurs in the vicinity of four circles
in the reciprocal (Fourier) space:
 \begin{eqnarray} 
 q_z = \pm \kappa _{||}\, ;\, |{\bf q}_\perp| = \kappa _\perp 
 .
 \label{lr7}
 \end{eqnarray}
Such a situation resembles that at the nematic - smectic $C$ phase transition \cite{SW76}.
The corresponding second order over ${\bm \varphi }$ terms in the free energy expansion 
in the Fourier space read as
\begin{eqnarray} 
\frac{1}{2}\sum _{\bf q}\left\{a + b(q_z, q_\perp) \right \}\bm\varphi ({\bf q}) \varphi (-{\bf q})
 ,
 \label{lr8}
 \end{eqnarray}
where $b(q_z, q_\perp )= b_{||}(|q_z|) - \kappa _{||})^2 + b_\perp (|{\bf q}_\perp | - \kappa _\perp )^2$.
The cross term $\propto (|q_z| - \kappa _{||})(|{\bf q}_\perp | - \kappa _\perp )$ can be excluded by an appropriate choice of the
coordinate frame basic vectors. Similarly, the 2-d and the 4-th order terms (\ref{lr3}), (\ref{lr4}), and  (\ref{lr6})
should be combined together.
\end{itemize} 
Ideologically theoretical descriptions of the all three possible $N_{TB}$ structures (easy-axis, easy-plane, and tilted)
are similar. Since experimentally identified $N_{TB}$ phase belongs to the easy-axis phase, in what follows (see the next subsection \ref{landau}) 
we analyze only easy axis $N_{TB}$ phase.

\section{Landau Theory of the easy axis $N_{TB}$ nematics}
\label{landau}
It is convenient to redefine the short wavelength order parameter ${\bm \varphi }$ to exclude 
its easy-axis modulation. Namely
we can introduce
\begin{equation}
 \bm\varphi= 2\,\mathrm{Re}\
 \left[\bm\psi \exp(i q_{||} z) \right].
 \label{bana3}
 \end{equation}
Entering this definition generally complex factor
${\bm \psi }$ is already long wavelength function. The vector ${\bm \psi }$ is orthogonal
to the director ${\bf n}$ and free energy expansions (\ref{bana1}) in terms of ${\bm \psi}$ has the following form
\begin{eqnarray}
 {\cal F}_\psi=\int dV \left\{a |\bm\psi|^2
 +b_1|\partial_\perp \bm\psi|^2
 +b_{||} |\partial_z \bm\psi|^2
 +b_2 |\nabla\cdot \bm\psi|^2
 \right. \nonumber \\ \left.
 +\frac{\lambda}{4} (\bm\psi \bm\psi^*)^2
 -\frac{\lambda_1}{4}
 \left[ (\bm\psi \bm\psi^*)^2
 -\bm\psi^2 (\bm\psi ^*)^2 \right]
 \right\} . \qquad
 \label{bana2}
 \end{eqnarray}
Free energy functional (\ref{bana2}) has more or less standard form of the Landau expansion for anisotropic
systems. It is easy to see that the last term in (\ref{bana2}) for $\lambda _1 >0$ gives the positive
contribution into ${\cal F}_\psi $. Therefore to minimize the free energy (\ref{bana2}) the term should be vanished.
There are two solutions vanishing this term
\begin{equation}
\psi _x = \pm i \psi _y
 \bm\varphi= 2\,\mathrm{Re}\
 \left[\bm\psi \exp(i q_{||} z) \right]
 \label{lr9}
 \end{equation}
and in the both cases ${\bm \psi }^2 = 0$.
Returning to the initial short wavelength order parameter ${\bm \varphi }$, the solutions (\ref{lr9}) can be rewritten as
\begin{equation}
 \varphi_x=2 |\psi_x|\cos(q_0z+\phi), \quad
 \varphi_y=\pm 2|\psi_x|\sin(q_0z+\phi),
 \label{cone}
 \end{equation}
where $\phi $ stands for an arbitrary phase of the complex short wavelength order parameter. 

To go beyond mean field theory, one has to include fluctuations. As usual in the vicinity of the $N$ - $N_{TB}$ phase
transition primary fluctuations are related to the short wavelength of the order parameter ${\bm \varphi}$. As
it is known \cite{BR75}, \cite{KLM93} the fluctuations of the short wavelength order parameter in the main over parameter  $a/(b_{||}q_{||}^2)$ 
approximation essential because of the large phase volume, where fluctuations are soft (i.e.,
easily excitable). However for the easy-axis $N_{TB}$ phase, the fluctuations are strong only around two points,
$q_z = \pm q_{||}$ in the reciprocal space. Therefore in this approximation (analogous to Hartree approximation
for electrons) the fluctuations of ${\bm \varphi }$ can be neglected.

However in the whole region of the existence of $N$ and $N_{TB}$ phases, there are also always soft fluctuations
of the long wavelength director ${\bf n}$. Taking into account that ${\bm \varphi}{\bm n} =0$,
and replacing $\bm n \to \bm n_0+\delta\bm n$, where $\delta\bm n$ is a small deviation of $\bm n$ from its average value. 
Then we arrive to the expressions 
 \begin{equation}
 \bm\varphi_\perp= 2\,\mathrm{Re}\
 \left[\bm\psi e^{i q_0 z} \right], \
 \varphi_\parallel= -\delta \bm n\cdot
 2\,\mathrm{Re}\
 \left[\bm\psi e^{i q_0 z} \right].
 \label{bana13}
 \end{equation}
instead of Eq. (\ref{bana3}). The subscripts $\parallel$ and $\perp$ signify the order parameter components parallel and perpendicular 
to the average director $\bm n$.

From (\ref{bana13}) and (\ref{bana1}) we derive the following interaction terms
 \begin{eqnarray}
 {\cal F}_{int1} = \int dV\
 b_2 \left[ i q_{||} \delta n_\alpha
 (\partial_\alpha \bm\psi \cdot \bm \psi^*
 \right. \nonumber \\ \left.
 -\bm \psi \cdot \partial_\alpha \bm \psi^*)
 +q_{||}^2 (\delta\bm n)^2 |\bm\psi|^2  \right],
 \label{bana14} \\
 {\cal F}_{int2} = \int dV\
 b_1 \left\{iq_{||} \left[
 (\delta\bm n\cdot \bm \psi) (\nabla\cdot \bm\psi^*)
 \right. \right.\nonumber \\ \left. \left.
 - (\delta\bm n\cdot \bm \psi^*)
 (\nabla\cdot \bm\psi) \right]
 +q_{||}^2 (\delta\bm n\cdot \bm \psi)
 (\delta\bm n\cdot \bm \psi^*) \right\}.
 \label{bana15}
 \end{eqnarray}
The terms (\ref{bana14},\ref{bana15}) can be obtained from  (\ref{bana2}) by replacing
$\partial_i \psi\to (\partial_i +iq_{||} \delta n_i)\psi$. This is a consequence of the rotational invariance of the system. The interaction terms should be added to the Landau functional (\ref{bana2}) and to the Frank energy.
 
To analyze the role of the long wavelength director fluctuations, one can apply well developed renormalization group (RG) procedure
(see, e.g., \cite{WK74}). As explicit RG calculations (following a schema presented in \cite{PP79} for the energy 
(\ref{bana14}) and (\ref{bana15})) prove (and common wisdom suggests) RG flow equations draw the system towards to
the symmetric fix point. It means that long wavelength ${\bf n}$ fluctuations renormalize the coefficient $b_1$  to zero.
In such a fixed point, the Free energy (\ref{bana1}) reads as 
\begin{eqnarray}
 {\cal F}_\psi=\int dV \left\{a |\bm\psi|^2
  +b_2 |(\nabla _\mu + i q_{||}\delta n_\mu) \cdot \bm\psi|^2
 \right. \nonumber \\ \left.
 +\frac{\lambda}{4} (\bm\psi \bm\psi^*)^2
 -\frac{\lambda_1}{4}
 \left[ (\bm\psi \bm\psi^*)^2
 -\bm\psi^2 (\bm\psi ^*)^2 \right]
 \right\} . \qquad
 \label{new22}
 \end{eqnarray}
The Free energy (\ref{new22}), supplemented by the Frank energy (\ref{Frank}), up to a self-evident change of notations (related to the vector nature of the order parameter
$\bm \psi $), coincide with the Free energy expansion for the nematic - smectic $A$ phase transition (see e.g., \cite{GP93},\cite{HL74},
\cite{LC78}).
Our model (\ref{new22}) is derived for the order parameter $\bm \psi $ with $n = 4$  in three-dimensional
space. As it was proved in \cite{LC78} in such a case a stable fixed point can be accessible only for
very large number of components of the order parameter ($n > 238$). Since for $n = 4$ there are no stable fixed points,
it is expected that there will be a first-order transition. This prediction is in agreement with experimental data
 \cite{PN10,CD11,KA12,ML13,MD14}. If this transition is a weak 1-st order phase transition,
fluctuation effects can be observed only in a narrow vicinity of the phase transition. Therefore the mean-field predictions 
hold in a broad temperature interval near the phase transition. For example small-angle X-ray diffraction studies \cite{BK13} 
demonstrate that the diffraction peak width at $T < T_c$ follows the mean-field prediction $\propto (T_c - T)^{1/2}$.

\section{Light scattering.}
\label {optics}

As it is well known, \cite{GP93}, \cite{LL84} intensity of light scattering is determined by fluctuations of dielectric permeability
$\delta \epsilon_{ik}$ for the system (or phase) under consideration.
These fluctuations are characterized by correlation functions of soft degrees of freedom, affecting $\delta \epsilon _{ik}$.
Dynamic correlation function ${\hat C}(\omega , {\bf q})$ for the dynamic light scattering ($\omega $ is frequency difference between the incident and scattered beams, and ${\bf q}$ is wave vector difference). Integrated over all frequencies $\int_{-\infty}^{+\infty} 
d \omega {\hat C}(\omega , {\bf q})$ yields to the static correlation function ${\hat C}({\bf q}$ and it determines the total scattered
intensity $I$. 
If one fixes the polarization of the incident light (${\bf i}$) and of the scattered light (${\bf f}$), the differential
scattering intensity $dI/d\omega d \Omega $ (per frequency $d\omega $ and solid angle $d\Omega $ intervals)
is  determined by the Rayleigh scattering law \cite{LL84}
\begin{eqnarray}
\frac{dI}{d\omega d\Omega } = \frac{\omega ^4}{32 \pi ^2} i_i i_m \langle \delta \epsilon _{ik}\, \delta \epsilon_{ms}\rangle f_k f_s 
. \qquad
 \label{lr22}
 \end{eqnarray}
There are various contributions to the thermal fluctuations of the dielectric permeability. The soft (i.e., easily excitable)
degrees of freedom dominates the scattering intensity.
In our model description of the $N$ and $N_{TB}$ structures, we exclude a thermodiffusion mode (assuming isothermic conditions),
sound modes (assuming incompressibility), and viscous shear modes (neglecting hydrodynamic motion). Then we end up with the following
degrees of freedom
\begin{itemize}
\item
In the $N$ phase these degrees of freedom are two components of the unit headless director ${\bf n}$. The corresponding
modes are hydrodynamic ones. Their eigen-frequencies $\omega $ vanish in the long wavelength limit (${\bf q} \to 0$).
\item
The sole genuine hydrodynamic mode in the $N_{TB}$ structure is related to the phase $\phi $ of the order parameter ${\bm \varphi }$
fluctuations. The matter is that at scales much larger than the pitch $q_{||}^{-1}$ of the heliconical structure in
the ordered state of the easy-axis $N_{TB}$, the module of the order parameter is frozen.
Then its Goldstone part (the phase $\phi $) variation in space is penalized 
by the following elastic energy
\begin{equation}
 {\cal F}_\mathrm{el}=\int dV\ \left[
 \frac{B_\perp}{2} (\partial_\perp\phi)^2
 +\frac{B_\parallel}{2}(\partial_z\phi)^2 \right],
 \label{super}
 \end{equation}
where the elastic moduli in the mean-field approximation are
\begin{equation}
B_\perp = 2(b_2 + b_1)|\psi _x|^2\, ;\, 
B_{||} = 4 b_{||}|\psi _x|^2 .
 \label{lr10}
 \end{equation}
As usually within Landau mean field approximation
\begin{equation}
|\psi _x| \propto {\sqrt {T_c - T}} ,
 \label{lr11}
 \end{equation} 
and therefore the scaling law for the elastic moduli
is
\begin{equation}
B_\perp\, ,\, B_{||} \propto (T_c - T)
 \label{lr12}
 \end{equation}
\end{itemize}
In the vicinity of the $N\, -\, N_{TB}$ phase transition point, the enumerated above hydrodynamic modes should be supplemented by
quasi-hydrodynamic ones, i.e., the modes which require a relatively small (but non-zero in the limit of ${\bf q}\to 0$) gap to
be excited. It is worth to noting that all modes we discus in this section are overdamped (diffusion-like) ones. Each of these non-propagating
modes describes relaxation of a single soft degree of freedom. Therefore the number of modes equals, which can contribute to
the light scattering spectra equals  to the number of soft degrees of freedom. Note to the point that it is not the case for
the propagating sound mode (excluded from our consideration due to incompressibility condition). Because of time inversion symmetry,
any propagating mode includes two degrees of freedom.

With all said above in mind
we arrive at the conclusion that in the vicinity of the $N\, -\, N_{TB}$ phase transition, in the $N$ state, one should
include into consideration five soft degrees of freedom. Namely, two Goldstone-like components of the director, and three
short-range correlated components of the
vector order parameter ${\bm \varphi }$. These five degrees of freedom corresponds to the five modes (two hydrodynamic and three pseudo-hydrodynamic ones) which can be detectable in the light scattering experiments. Similarly from the $N_{TB}$ side
there are six soft degrees of freedom, forming six modes potentially observable in the light scattering experiments.
Namely, one hydrodynamic mode, describing phase of the long ranged order parameter fluctuations, and five 
quasi-hydrodynamic modes. The matter is that in the $N_{TB}$ state the director orientation is coupled 
to the vector order parameter ${\bm \varphi}$. Thus the director and ${\bm \varphi}$ orientation
fluctuations acquire an energy gap $\omega \neq 0$ at ${\bf q} \to 0$.
Nevertheless in the vicinity of the $N$ -  $N_{TB}$ phase transition these modes are not too fast (the gap in
the excitation spectra is relatively small), therefore the modes can be detected in the light scattering experiments 
see \cite{PS16}, where light scattering experimentally data are rationalized in terms of a slightly different from presented
here model of the $N_{TB}$ liquid crystals.
Especially delicate issue is an observation of the new hydrodynamic (Goldstone) mode in the $N_{TB}$
structure (related to long-scale variations of the order parameter ${\bm \varphi}$ phase $\phi $. Identification of this mode by optical scattering methods could requires a very accurate selection of polarizations for the incident and scattered beams polarized to exclude presumably much larger scattering by conventional director modes. Second problem is that the optical wave vector is smaller than the inverse 
pitch of the $N_{TB}$ heliconical structure. Then only second order scattering 
processes (proportional to the square of the $N_{TB}$ order parameter fluctuations) contribute to the light scattering intensity. 
To the point, an external magnetic (or electric) field which suppresses the conventional long-scale director fluctuations can be very useful for the observation of the mode (see also subsection \ref{electric}).

\section{Rheology of the $N_{TB}$ nematics.}
\label{rheology}

Results presented in this section \ref{rheology} is motivated by two  
recent works \cite{KK20} - \cite{KK21} on rheological studies of the $N_{TB}$ liquid crystals.
In what follows we integrate the input from these works, supplemented
by my own contribution to this field \cite{KA22}.
In an abbreviated form, the results of rheological experiments (stationary shear $\dot {\gamma }$ 
of the ordered $N_{TB}$ liquid crystal)
presented in the papers can be summarized as follows. There are three well separated regions of 
the applied shear:

\begin{enumerate}

\item
In the region $I$ for ${\dot \gamma } < {\dot \gamma }_{c1}$ the shear stress $\sigma $ scales
as $({\dot {\gamma }})^{1/2}$. According to the definition $\sigma \propto \eta _{eff} {\dot \gamma }$
in such conditions the defined above effective viscosity $\eta _{eff}$ decreases under shearing as 
$\eta _{eff} \propto {\dot \gamma }^{-1/2}$. This is so-called shear-thinning behavior.
The proportionality coefficient $C_1$ depends on temperature $\propto (T_* - T)^{1/2}$ where $T_*$ is slightly below
$N\, -\, N_{TB}$ transition temperature $T_c$ (for the material studied in the works \cite{KK20} - \cite{KK21} 
$T_* \simeq 103 ^{\circ} \, C$ and $T_c \simeq 108 ^{\circ} \, C$).  
\item
In the region $II$ when ${\dot \gamma }_{c1} < {\dot \gamma } < {\dot \gamma }_{c2}$
there is a sort of plateau, i.e., $\sigma $ is independent of ${\dot \gamma }$.
Hence $\eta _{eff} \propto {\dot \gamma }^{-1}$, i.e., again (like in the region $I$) shear-thinning
rheology.
\item
For a relatively high applied shear stress, region $III$, ${\dot \gamma } > {\dot \gamma }_{c2}$
the standard Newtonian rheology takes place, when $\sigma $ scales as ${\dot \gamma }$, and $\eta _{eff}$
is shear independent.
\item
There is a jump in the value of  $\sigma $ at the transition from the region $II$ to region $III$,
and $\Delta \sigma $ decreases with $T$ and vanishes at $T_*$.

\end{enumerate}

Below in this section we will present heuristic theoretical interpretation of these results.
The underlying idea is that the shear tends to suppress the thermal
fluctuations of the layers, decreasing the entropy of the system and, hence,
increasing its free energy. This effect depends on the orientation of the layers
with respect to the velocity and the velocity gradient.

Our interpretation is based on the coarse grained (i.e., obtained by averaging over scales much
larger than the $N_{TB}$ heliconical structure pitch) description of the $N_{TB}$ phase.
As it was already said., structurally the $N_{TB}$ structure (although not possessing any density
modulation) at scales larger than the pitch of the heliconical  director $\bm n+\bm\varphi$ is equivalent
to the smectic $A$ liquid crystal. In such a description periodically modulated orientation
of the heliconical state can be described by the equidistant layers equivalent to the smectic layers.
Therefore the $N_{TB}$ coarse grained elastic energy 
reads as
\begin{equation}
{\tilde {F}} = \frac{1}{2}\int d^3 r\left [ {\tilde B}\left (\frac{\partial u}{\partial z}\right )^2 + {\tilde K}\left 
(\nabla _\perp ^2 u\right )^2\right ]
.
 \label{kats3}
 \end{equation}
Here $u$ is a layer displacement along the normal to the layer ($z$ axis in our notation), and coarse grained
elastic moduli \cite{GP93}, \cite{KL93} are 
\begin{equation}
{\tilde B} = b_2 q_{||}\, ;\, {\tilde K} = \frac{3}{8} b_2
.
 \label{kats4}
 \end{equation}

One more note is in order here. There are known in the literature other coarse grained descriptions of the $N_{TB}$
phase \cite{SK14}, \cite{PS16}, \cite{MD16}. Although the approaches \cite{SK14}, \cite{PS16}, \cite{MD16} are conceptually similar 
to the model presented in this section, one essential difference should be emphasized. 
In our approach the long wavelength (nematic director) and
short wavelength the $N_{TB}$ order parameter are explicitly separated within the framework of the Landau free energy expansion (\ref{bana1}). Technically the short wavelength nature of the $N_{TB}$ order parameter yields to the specific form of the gradient terms
which provide a softening of the order parameter at a finite wave vector $q_{||}$ (not around at ${\bf q} =0 $ 
as it is the case for the nematic order parameter). 
In the $N_{TB}$ phase but not too far from the transition
point $T_c$ (that is always the case due to $(T - T_c)/T_c \ll 1$) all phenomenological coefficients
(except $a$) entering the mean field Landau model (\ref{bana1}) can be considered as temperature independent.
The sole controlling mean-field behavior coefficient $a$ scales
as $T - T_c$. 

Similar coarse-grainig should be performed
also to express smectic-like 5 viscosity coefficients \cite{KL86} in terms of the bare (nematic-like)
According to the definitions \cite{MP70}, \cite{MP72}, \cite{GP93}, \cite{KL93} the uniaxial nematic phase $N$ is characterized
5 independent viscosity coefficients $\eta _1 \, -\, \eta _5$ (or 6 Leslie's
viscosity coefficients, which have to satisfied one constraint \cite{GP93}). 
At small space scales (on the order of the pitch) the $N$ and $N_{TB}$ viscosity coefficients can be different, however the 
difference is relatively small 
(on the order of $\bm\varphi ^2$). We neglect the difference in what follows. For two rheological configurations (see figure 1), 
which allow hydrodynamic flow not hindered by the layer structure, only two combinations of the five viscosity 
coefficients needed for a full description of dynamical behavior.
The flow is determined by these two combinations of the coarse grained viscosity coefficients.
Namely, ${\tilde \eta }_2$ coefficient determines the shear flow rheology for the perpendicular configuration,
and the coarse grained viscosity ${\tilde
\eta }_3$ determines the flow in the parallel configuration.

It is worth to stress that coarse grained
free energy of the $N_{TB}$ phase coincides with the smectic $A$ free energy, provided the smectic layer displacement $u$
is replaced by the phase of the $N_{TB}$ order parameter.
Similar statements holds for the coarse grained dynamics of the $N_{TB}$ phase. However we have to keep in mind that 
the coarse grained viscosity coefficients have a physical meaning only at the scales $r$ larger than $q_{||}^{-1}$. At smaller
scales rheological behavior is determined by the bare (nematic) viscosity coefficients. Therefore when the shearing liquid crystal sample
thickness becomes smaller than $q_{||}^{-1}$ the coarse grained approximation is meaningless. In particularly no any room for the coarse
grained theory in the $N$ phase, where the $N_{TB}$ order parameter $|\bm \varphi | \to 0$ and hence $q_{||} \equiv 0$.

Due to the anisotropy of the $N_{TB}$ phase there are five independent viscosity coefficients, entering the forth order viscosity tensor 
$\eta_{ijkl}$ \cite{GP93}. The viscous dissipative stress tensor at small scales reads as
\begin{eqnarray}
{\hat {\sigma}_{eff} }=
\eta_{ijkl} A_{kl} = 2\eta _2 A_{ij}
+ \, 2(\eta_3 - \eta_2)(A_{ik}n_kn_j + A_{jk}n_i n_k)
\nonumber \\
\label{viscosity1}
 +(\eta_4 - \eta _2)\delta_{ij}A_{kk} +2(\eta_1 + \eta_2 - 2\eta_3) n_in_jn_kn_lA_{kl}
\nonumber \\  +
(\eta_5 - \eta_4 + \eta _2)(\delta _{ij}n_k n_lA_{kl} + n_i n_j A_{kk}) \, ,
\nonumber
\end{eqnarray}
where $A_{ij} =\partial _i v_j + \partial _j v_i$.

\begin{figure}
\begin{center}
\includegraphics[height=2in]{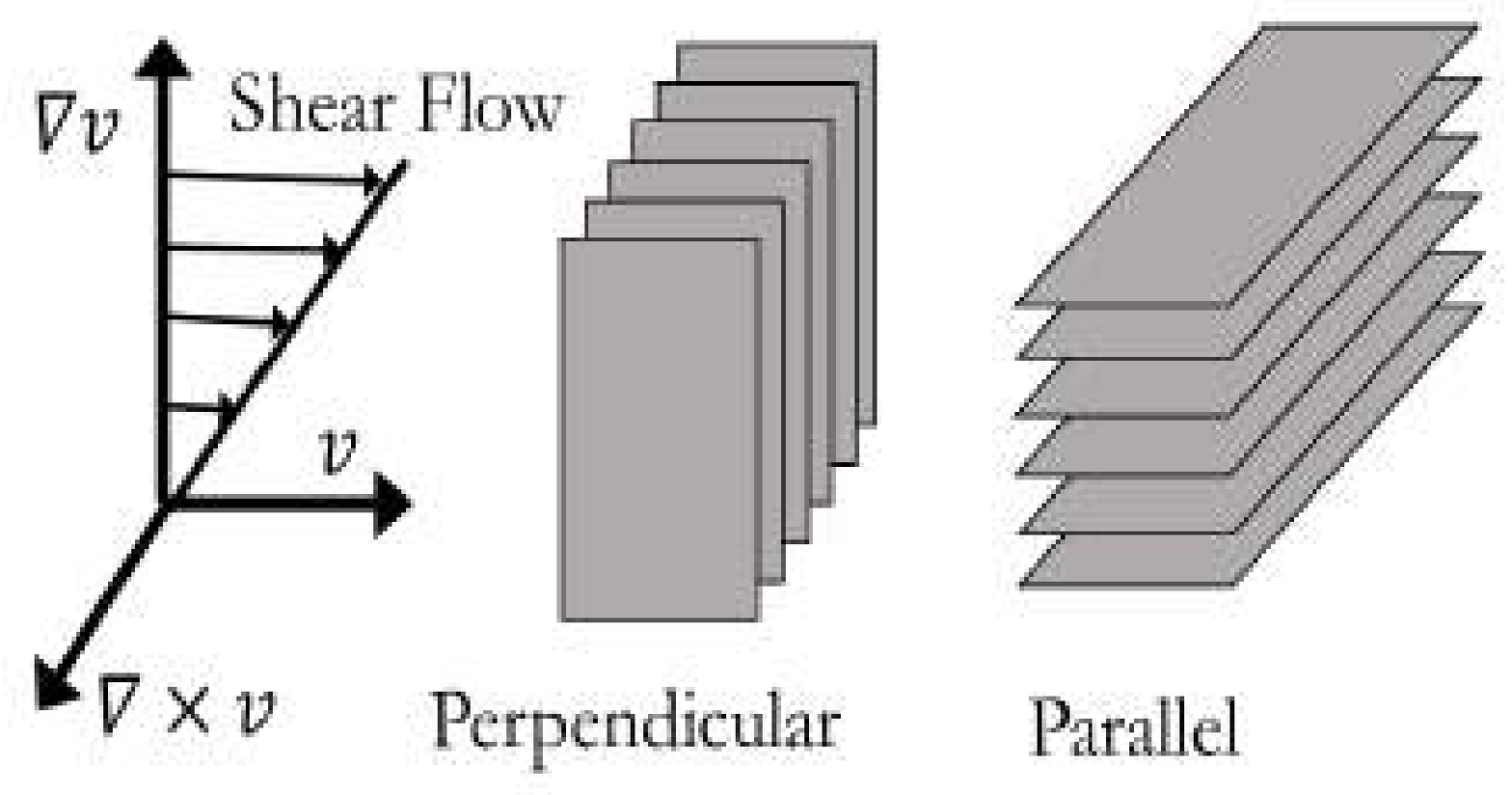}
\end{center}
\caption{Two rheological configurations considered in the paper.}
\label{f1}
\end{figure}

In the rheological configurations shown in figure \ref{f1} shear flow dissipation is 
determined by the two coarse grained viscosity coefficients 
${\tilde \eta}_2$ and ${\tilde \eta}_3$. These coefficients
can be expressed \cite{KL86} in terms of the bare nematic viscosity coefficients
($\eta _i$, with $i = 1,...5$) as
\begin{equation}
{\tilde \eta}_2 = 2\eta _2 - \frac{1}{2} \eta _3 + \frac{1}{4}\eta _1\, ;\, 
{\tilde \eta}_3 = \frac{1}{4}\left [2\eta _3 + 2\eta _2 - \eta _1\right ]
.
 \label{kats5}
 \end{equation}
For typical in nematic liquid crystals values of the bare viscosity
coefficients \cite{GP93}, \cite{KL06}, \cite{OP06}, we estimate from (\ref{kats5})
the values of the needed for us coarse grained
viscosity coefficients. Namely in the limit of shear rate ${\dot \gamma } \to 0$: 
\begin{equation}
{\tilde \eta}_2^0 \simeq 0.832\, Poise\, ;\, 
{\tilde \eta}_3^0 \simeq 0.2\, Poise
.
 \label{kats6}
 \end{equation}

Because in the limit ${\dot{\gamma}} \to 0 $, ${\tilde \eta }_2 > {\tilde \eta}_3$ just the perpendicular configuration $a$ is preferable
one (the configuration leads to the maximum rate for the entropy production). 
This principle (the maximum entropy production rate) as well as the apparently controversial statement (the principle of minimum entropy production principle) in each particular condition should be reconsidered, based on statistical mechanics,
and hydrodynamics to find a stationary state for non-equilibrium systems. In this work we rely on the mapping of 
coarse grained dynamics of the $N_{TB}$ phase into that for the smectic $A$ liquid crystals.
With this mapping in hands the three regimes observed in the sheared $N_{TB}$ liquid crystals can be rationalized 
similarly to the known for smectic $A$ results (see e.g., two review articles \cite{SB10}, \cite{FK14}), and the results
suggest three rheological regimes which are governed by the maximum entropy production rate (irrespective to physical
mechanisms behind). 

The main feature which distinguishes the standard nematic $N$ and the twist-bend nematic $N_{TB}$ liquid 
crystals is a short
wavelength modulation of the orientation order ${\bm \varphi }$ presented in the $N_{TB}$ phase.
This two component vector ${\bm \varphi }$, orthogonal to the nematic director ${\bf n}$ (${\bm \varphi} \cdot  {\bf n}
= 0$) can be chosen as the order parameter describing $N$ - $N_{TB}$ phase transition. 
Since at the scales $r$ larger than $q_{||}^{-1}$ we are interested in this work, the $N_{TB}$ phase free energy
and dynamic equations coincide with those for the effective smectic $A$, in what follows we will use interchangeably the both terms. 
$N_{TB}$ nematics when discussing generic features of the $N \to N_{TB}$ phase transition, and effective smectic $A$, speaking about 
rheology. 
It is worth to emphasize, that there is no density modulation in the liquid-like
$N_{TB}$ phase. However, conical $N_{TB}$ orientation order, on scales much larger than the orientation 
modulation period looks like as a periodic in space layered smectic-like structure.

A wide range of coarse-grained models have been proposed, usually dedicated to modeling of multiscale systems.
Coarse graining allows to decrease a number of essential degrees of freedom at the expense of microscopic details
(e.g, replacing individual building blocks by their groups). Technically in this work we use slightly modified to be applicable
to the $N_{TB}$ phase coarse-graining procedure described in \cite{GP93} for the static behavior of cholesterics,
and in \cite{KL86} - for dynamics of cholesteric liquid crystals. 
The properties of the deformed $N_{TB}$ liquid crystal depends essentially on the ratio of the inhomogeneity
scale (say $r_{in}$) and the $N_{TB}$ heliconical spiral pitch $q_{||}^{-1}$. At scales $r_{in} \gg q_{||}^{-1}$ 
to find large-scale static and dynamic characteristics, we have to eliminate fast degrees of freedom. It is done
\cite{GP93}, \cite{KL86}
by appropriate integration out the fast degrees of freedom. Performing the averaging we express the coarse grained
coefficients (elastic moduli and viscosity coefficients) in terms of the bare $N_{TB}$ parameters defined at scales
$\ll q_{||}^{-1}$.

Some of the points of the derivation presented below, can be found in the work \cite{KA22}.
Here we present a much simpler derivation which offers a deeper insight into the physics behind the $N_{TB}$ 
rheology. We will use a scaling-like approach. Even though the approach may overlook some details, the obtained scaling
laws provide a quick picture of the rheological behavior in terms of a few appropriate parameters, and how essential
quantities vary as functions of these parameters.

Thus we assume the ideally ordered initial configuration. Shear stress and strain in this configuration
is determined by the coarse grained viscosity ${\tilde {\eta }_2}$.
External shear produces certain stress in the flowing $N_{TB}$ material. 
We estimate this effective stress following the arguments developed for shearing smectic $A$
liquid crystals in the works \cite{SB10}, \cite{FK14}, \cite{PA95}. 
The stress tensor in the $N_{TB}$ (or in any other, e.g., smectic or cholesteric, effectively lamellar structure)
can be estimated as
\begin{equation}
\sigma _{eff} \simeq \frac{\sqrt{B K}}{R_d}
,
 \label{rh1}
 \end{equation}
where $B$ and $K$ are large scale (coarse grained) elastic moduli, and $R_d$ is characteristic scale of the
produced by external shear deformation. The scale $R_d$ depends on the shear setup geometry. For the ideally ordered
$N_{TB}$ sample, there are possible only two configurations which allow the shear flow. The configurations will be termed
in what follows as orthogonal ($a$ configuration) and parallel ($c$ configuration). The configurations are schematically shown
in the figure \ref{f1}. 

In the orthogonal $a$ configuration the normal to the layers, ${\bf l}$ is orthogonal to the shear gradient $\nabla {\bf v}$,
whereas in the parallel configuration $c$ it is parallel to $\nabla {\bf v}$.
It is worth to noting the striking differences with conventional fluids or nematic liquid crystals which comes from the existence of 
a new length scale in the lamellar liquid crystals. In the lamellar liquid crystals when  ${\bf v}$ is 
along ''solid-like'' direction (${\bf v} || {\bf l}$) the
shear (Poiseuile like) flow is impossible. In such configuration (termed in the literature \cite{GP93} as $b$ configuration),
so-called plug flow occurs. The latter one is characterized by the penetration length
\begin{equation}
\delta _p = (\lambda_p {\tilde \eta }_3^0)^{1/2}
,
 \label{rh2}
 \end{equation}
where $\lambda _p$ is the permeation coefficient. Typically in lamellar liquid crystals
$\lambda _p \simeq 10^{-8} \, -\, 10^{-9}\, cm^3 s/g$, and then the penetration
length $\delta _p \simeq 0.5\, \mu m$.

At large scales, we are interested in this section, the bare values of the viscosity coefficients should be
replaced by their coarse-grained values (see, \cite{KL86}, and (\ref{kats5}).
As a note of caution it is worth to noting that the maximum entropy production rate principal (as well as the apparently controversial statement, the principle of minimum entropy production principle)in each particular condition should be reconsidered, based on statistical mechanics,
and hydrodynamics to find a stationary state for non-equilibrium systems. In this work we rely on the mapping of 
coarse grained dynamics of the $N_{TB}$ phase into that for the smectic $A$ liquid crystals.
With this mapping in hands the three regimes observed in the sheared $N_{TB}$ liquid crystals can be rationalized 
similarly to the known for smectic $A$ results (see e.g., two review articles \cite{SB10}, \cite{FK14}), and the results
suggest three rheological regimes which are governed by the maximum entropy production rate (irrespective to physical
mechanisms behind). 
Because in the limit ${\dot{\gamma}} \to 0 $, ${\tilde \eta }_2 > {\tilde \eta}_3$ just the perpendicular $a$ configuration is preferable
one (the configuration leads to the maximum rate for the entropy production). We start with the $a$ rheological experiment geometry.

In the $a$ orientation external shear flow produces non-uniform in space deformation energy along
$\nabla {\bf v}$ direction (see figure \ref{f1}. This direction $\nabla {\bf v}$ is the only non-uniformity direction
for the shearing $N_{TB}$ structure in the $a$ configuration. The characteristic space scale along this
direction can be estimated by comparing viscous and elastic forces.
The latter one is estimated by the Frank elastic modulus $K$, whereas the former one is estimated by the Stokes viscous
force
\begin{equation}
{\tilde {\eta }}_2^{(0)} R_d^{(a)} v \simeq {\tilde {\eta }}_2^{(0)} R_d^{(a)}
{\dot {\gamma }} 
,
 \label{rh3}
 \end{equation}
where superscript $(a)$ stands for the $(a)$ shear rheological configuration, and ${\tilde {\eta}_2^{(0)}}$
is ${\tilde {\eta _2}}$ at ${\dot {\gamma }} \to 0$.
Therefore 
\begin{equation}
R_d^{(a)} = {\sqrt{\frac{K}{{\tilde \eta}_2^{(0)}}}}({\dot {\gamma }})^{-1/2}
,
 \label{rh4}
 \end{equation}
Combining this relation with the expression for the stress tensor (\ref{rh1}) we arrive
to the following rheological relation in the parallel configuration $a$
\begin{equation}
{\hat {\sigma }_{a}} = {\sqrt {B {\tilde {\eta }}_2^{(0)}}}{\dot {\gamma }^{1/2}}
.
 \label{rh5}
 \end{equation}
We conclude from (\ref{rh5}) that in the geometry $a$ we have deal with non-Newtonian 
rheology (shear thinning), because from (\ref{rh5}) follows that
\begin{equation}
{\tilde {\eta }}_2^{(0)}({\dot {\gamma} \neq 0}) \propto A_1(T){\dot {\gamma }^{-1/2}}
.
 \label{rh6}
 \end{equation}

Since in the mean field approximation ${\tilde {B}}$ vanishes near the $N_{TB} \, -\, N$ transition point as
\begin{equation}
{\tilde {B}} \propto (T_* - T)
,
 \label{rh7}
 \end{equation}
where $T_*$ is close to the $N_{TB} \, -\, N$ transition point (although does not coincide with it for the first
order phase transition). Then the prefactor $A_1(T)$ in (\ref{rh6}) has the following scaling
\begin{equation}
A_1(T) \propto (T_* -T)^{1/2}
.
 \label{rh8}
 \end{equation}

In the rheological $c$ configuration the stress tensor is estimated similarly as that in the $a$ configuration (\ref{rh1}).
However the deformation force in the $c$ orientation should be defined differently. The matter is that in the $c$ configuration,
the imposed by shear non-uniformity direction ($\nabla {\bf v}$) is along the director (normal to the layers). Therefore
the elastic force is determined like in the Helfrich undulation instability in smectics or cholesterics \cite{GP93}, \cite{KL06},
\cite{OP06}. Namely 
\begin{equation}
f_{el} \simeq \sqrt{KB} d
.
 \label{rhnew1}
 \end{equation}

The elastic force (\ref{rhnew1}) equilibrates the Stokes friction force

\begin{equation}
f_{vis} \simeq \eta_3^0 {\dot {\gamma }} d R_d
.
 \label{rhnew2}
 \end{equation}
then
\begin{equation}
R_d = \frac{{\sqrt{KB}}{\eta_3^0{\dot{\gamma }}}} d
,
 \label{rhnew3}
 \end{equation}
and we end up with
\begin{equation}
{\hat {\sigma }_{c}} = {\tilde {\eta }}_3^{(0)}{\dot {\gamma }}
.
 \label{rh15}
 \end{equation}
Thus, in the $c$ configuration we recover Newtonian rheological behavior with $\sigma_c \simeq \eta_3^0 {\dot{\gamma}}$.

As we mentioned already above the maximum entropy production rate principle suggests $a$ configuration at small ${{\dot \gamma }}$.
However, due to shear thinning phenomenon upon increasing ${\dot \gamma }$, the rheological $c$ configuration becomes favorable.
In the shearing $N_{TB}$ state transformation $a \to c$ configuration can occur by gradual rotation of the $a$ domains.
In the process of the transformation unavoidably appears domains where ${\bf v}$ is 
along ''solid-like'' direction (${\bf v} || {\bf l}$). The stress tensor ${\hat {\sigma }_{mix}}$ in such mixed state
is independent of the shear rate and can be estimated as
\begin{equation}
{\hat {\sigma }_{mix}} \simeq B
.
 \label{rh11}
 \end{equation}
Eq. (\ref{rh11}) can be also rationalize as a natural sign of no-shear flow condition at relatively small ${\dot{\gamma }}$ along the normal to the layers.

With obtained above expressions and estimations for the stress tensors in three rheological regimes (${\hat \sigma}_a$, ${\hat \sigma}_{mix}$,
${\hat \sigma}_c$)  in hands we are in the position to find all essential physical characteristics of  the rheological phase diagram.
Its topology (i.e., relative positions of the ${\hat \sigma}_a$, ${\hat \sigma}_c$,
and ${\hat \sigma}_c$ branches) is chosen to confront our results with experimental data \cite{KK20}, \cite{KK21}.
Namely, the first critical value of the shear rate ${\dot{\gamma}_{c1}}$ can be found from the condition
${\hat{\sigma}_a(\dot{\gamma}_{c1})} = {\hat{\sigma}_{mix}}$.
Thus from (\ref{rh5}) and (\ref{rh11}) we find
\begin{equation}
{\dot{\gamma}_{c1}} \simeq \frac{B}{\eta _{2}^{(0)}}
.
 \label{rh111}
 \end{equation}

Similarly the second critical value of the shear rate ${\dot{\gamma}_{c2}}$
can be found from the condition ${\hat{\sigma}_c({\dot{\gamma}_{c2})}} =  
{\hat{\sigma}_a({\dot{\gamma}_{c2})}}$, which according to Eq. (\ref{rh5}) and (\ref{rh15})
reads as  

\begin{equation}
{\dot{\gamma}_{c2}} \simeq \frac{B\eta _{2}^{(0)}}{(\eta_3^{(0)})^2}
.
 \label{rh112}
 \end{equation}
Finally the jump $\Delta \sigma^*$ is determined as
\begin{equation}
{\hat{\sigma}_a({\dot{\gamma}_{c2}})} - {\hat{\sigma}_{mix}} \simeq B\left\{\left[
\frac{\eta _{2}^{(0)}}{\eta_3^{(0)}}\right]^2 - 1\right\}
. 
 \label{rh113}
 \end{equation}

The calculated rheological phase diagram reproduces reasonably well not only observed experimentally shear thinning phenomena
in the $N_{TB}$ state. Equally well it fits experimental data concerning temperature dependencies of the parameters entering the rheological laws.
Let us recall once again the main experimental results \cite{KK20}, \cite{KK21}.
The authors of these papers found nontrivial non-Newtonian behavior
of sheared $N_{TB}$ nematics. At relatively low shear rate (${\dot \gamma } \leq {\dot \gamma}_{c1}$) the stress
tensor $\sigma $ created by this shear strain, scales as $\sigma \propto C_1(T) {\dot \gamma }^{1/2}$ (where the coefficient $C_1(T)$ is independent of ${\dot{\gamma }}$). The effective viscosity decreases with the shear rate ($\eta \propto {\dot \gamma }^{-1/2}$) 
manifesting so-called shear-thinning phenomenon. At intermediate shear rate ${\dot \gamma }_{c1} \leq {\dot \gamma} \leq {\dot \gamma }_{c2}$, 
${\hat{\sigma}_{mix}} $ 
is almost independent of ${\dot \gamma }$ (a sort of plateau), and at larger shear rate (${\dot \gamma } \geq 
{\dot \gamma}_{c2}$), ${\hat{\sigma}_c} \propto {\dot \gamma }$, and it looks like Newtonian rheology.
Measured experimentally coefficient $C_1(T)$ vanishes as $\propto (T - T^*)^{1/2}$ at a certain temperature $T^*$ (below $N\, - \, N_{TB}$ 
phase transition point $T_c$, where $N$
stands for conventional nematic state). 
For the stress tensor ${\hat{\sigma}_c}$ experimental data suggest no critical-like temperature
dependence and a jump at 
${\dot \gamma} = {\dot {\gamma }}_{c2}$. The jump decreases with $T$ and
vanishes at the temperature $T^*$.
Found in our work expressions for ${\hat{\sigma}_a}$, ${\hat{\sigma}_{mix}}$, and ${\hat{\sigma}_c}$ predict 
that ${\hat{\sigma}_{a}}$ contains the factor $\sqrt B$ and therefore indeed scales as $(T^* - T)^{1/2}$ in the agreement
with experimental data.

Thus in the agreement with experimental data $\Delta \sigma^*$ is positive
and decreases with temperature like $(T_* - T)$.

Although all details of the observed experimentally rheological behavior of the $N_{TB}$ nematics still remains to be clarified,
the main message of this section is robust. Namely we claim that coarse grained dynamic description of the $N_{TB}$ phase is
allows to rationalize qualitatively the observed in the phase different rheological regimes. Such coarse grained description supplemented by
arguments based on the maximum rate of entropy production principle, can be used to estimate
the critical values of the shear rate separating these rheological regimes. The estimations made are applicable
to the shearing of well ordered samples of the $N_{TB}$ phase. If it is not the case, defects (disclinations or domain walls)
affect the rheology. However relying on the maximum entropy production rate principle, we expect that the rheological transitions
(crossover between different rheological regimes) are determined by the coarse grained effective viscosity coefficients. The latter ones
should be determined for the partially disordered (i.e., including defects) $N_{TB}$ phase flow.

The main message of this section is that coarse grained dynamic description of the $N_{TB}$ phase
allows to rationalized observed in the phase different rheological regimes. Such coarse grained description supplemented by
qualitative arguments based on the maximum rate of entropy production principle, can be used to estimate
the critical values of the shear rate separating these rheological regimes. The estimations made in the paper are applicable
to the shearing of well ordered samples of the $N_{TB}$ phase. If it is not the case, defects (disclinations or domain walls)
affect the rheology. However relying on the maximum entropy production rate principle, we expect that the rheological transitions
(crossover between different rheological regimes) are determined by the coarse grained effective viscosity coefficients. The latter ones
should be determined for the partially disordered (i.e., including defects) $N_{TB}$ phase flow. 
Guided by the modest aim of this work, we present only scaling analysis of  three regimes of a steady shear  viscosity
curve which is in qualitative agreement with previously reported observations \cite{KK20}, \cite{KK21}, \cite{PS16}.
Even more not all of the observed qualitative features of the rheological curves are reproduced in our approach.
For example there is no independent of shear rate plateau in the interval ${\dot \gamma }_{c1} \leq {\dot \gamma} \leq {\dot \gamma }_{c2}$.
Instead of that $\sigma \propto {\dot \gamma }^{\alpha }$, where the exponent $\alpha \simeq 0.1$ is small but not zero.
The matter is that the estimated thresholds of rheological curves assume a
well-defined orientation of the director respective to the shear flow. Most probably it is not the case
in mentioned above experimental works, especially in the regions of relatively small shear rates. However, 
the ordered, defects-free states can be achieved, e.g., by orienting liquid crystal external
field. Such investigation is beyond the scope of this work, direct experimental measurements of the ordered $N_{TB}$ phase
rheology is still a challenging task. I hope that the results of this section will stimulate
discussions on the intriguing and important issues of non-Newtonian rheology in liquid crystals.

\section{External field effects.}
\label{electric}

As it is well known (see any textbook on general thermodynamics, e.g., \cite{LL80},  \cite{ST87}, \cite{HU87},
\cite{CL00}, or its specially adapted for liquid crystals versions in 
\cite{PI91}, \cite{AN91}
\cite{CH92}, \cite{BC94}, \cite{KL06}, \cite{OP05})
for the first order phase transitions, the shift of the
transition temperature in an external (magnetic or electric) field is 
related to the susceptibility $\Delta \chi $ difference (in what follows for the sake of simplicity we focus on only magnetic field) in the ordered and disordered
states. In the case of a quadrupole ordering (like in nematic liquid crystals) this shift of the transition temperature $\Delta T_c$ is expected to be quadratic over the field $H$ ($\Delta T_c \propto H^2$). For typical nematic material parameters 
\cite{CH92}, \cite{GP93},
\cite{BL11}, \cite{OP05}, $\Delta T_c$ is very small (on the order of a few $mK$ for $H \simeq 10^6\, G$). In the frame work of the
Landau - de Gennes mean field theory $\Delta T_c(H)$ (see the original paper \cite{HE70} or textbooks \cite{CH92},
\cite{GP93}) reads as 
  \begin{equation}
    \label{ntb1}
    \Delta T_c(H) \propto \frac{\Delta \chi H^2}{|L|}
		    \, ,
  \end{equation}
  where 
$L$ is the latent heat of the 1-st order isotropic liquid - nematic liquid crystal ($I - N$) transition. In terms of the Landau expansion over the order parameter $Q_{ik} =
(3/2)S(n_in_k - (1/3)\delta _{ik}$ for the uniaxial orientation order (where $S$ is  the module of the order parameter, and
${\bf n}$ is director, ${\bf n}^2 = 1$):
 \begin{equation}
    \label{ntb2}
    F = \frac{1}{2} a Q_{ik}^2 - \frac{1}{3} \mu Q_{ik}^3 + \frac{1}{4} \lambda_4 Q_{ik}^4 - \frac{1}{2} \Delta \chi Q_{ik} H_i H_k
		    \, ,
  \end{equation}
	where as usual $a = \alpha (T - T_*)$ ($T_*$ is a temperature where the isotropic phase becomes  unstable, and for a weak 1-st
	order phase transition $T_*$ is close to the transition temperature $T_c$), scaling Exp. (\ref{ntb1}) has the following explicit form
   \begin{equation}
    \label{ntb3}
    \Delta T_c(H) = 3\frac{\lambda_4}{\mu}\frac{\Delta \chi  H^2}{\alpha }
		    \, .
  \end{equation}
  This formula leads to a mentioned above mean field prediction of a very small (for any realistic magnetic field) and quadratic over
	the field shift of the transition temperature
	for the  $I - N$ phase transition.
This was a state of the art in this business when (see e.g., \cite{HS09} - \cite{MD14})
a new kind of liquid crystalline ordering, so-called	twist-bend nematics have been discovered.
As it is often the case, the discovery of new phases motivates new researches to elucidate their properties. Then it became clear that even conventional uniform nematic phase of such compounds reveals unusual physical properties. For instance the authors of the works \cite{ST16}, 
and also \cite{OW08}, \cite{FV11}, \cite{TS13} have observed  
for dimer-shape mesogenes and in particularly for some bent-core molecules (we will use the both terms, bent-core and dimer interchangeably) the anomalously large $\Delta T_c(H)$ up to $15\, K$ in about $20\, T$ magnetic field. What is even more 
surprising, is that the transition temperature shift increases almost proportionally to $H$, and is not proportional to $H^2$.

The authors of the experimental observation \cite{ST16} attributed the unprecedentedly large $\Delta T_c$ to a field-induced straightening 
of the dimer (or bent-core) mesogens conformation. In own turn the microscopic straightening of the mesogens in macroscopic language should be translated to an enhance of the modulus $S$ of the order parameter. In the mean-field approximation just this enhance leads to $H^2$ scaling
for the $\Delta T_c$. Therefore to explain the unusual scaling $\Delta T_c(H)$ one has to go beyond the mean-field approximation. Indeed as it is well-known nematic liquid crystal - isotropic liquid phase transition is typically a weak first-order phase transition with pronounced fluctuation effects \cite{AN91}. In the critical region (typically a few $K$ around the transition point) the fluctuations of the  order parameters are described by 5 independent degrees of freedom or modes (see e.g., \cite{PK77}) 
\begin{itemize}
\item
2 gapless transverse director modes, with their dispersion law $\omega \propto i q^2$;
\item
two biaxial modes with the gap proportional $\mu$ (where as above $\mu$ is the coefficient in the third order term in the Landau expansion
(\ref{ntb2}));
\item
a single mode related to the modulus $S$ of the order parameter fluctuation with the gap $\propto a$, which can be zero only at the spinodal
temperature $T_*$ (as usual in the Landau theory  (\ref{ntb2}) $a \propto (T - T_*)$).
\end{itemize}
However, as it was shown for nematic liquid crystals in \cite{PK77}, there is another indirect source of the modulus $S$ fluctuations - so-called singular longitudinal fluctuations. The fluctuations are induced by the soft (Goldstone)
transverse fluctuations of the director - therefore can be essential not only in the critical region near the transition point but in the 
entire range of the nematic phase stability.

Our main motivation in this section is to investigate the role of these singular longitudinal fluctuations of the order parameter as a macroscopic counterpart to the microscopic mechanism of the straightening dimer mesogens
conformations, proposed in \cite{ST16}. In the next subsection \ref{bas}
we derive basic relations needed to include longitudinal fluctuations of the nematic order parameter, and give a short,
very incomplete summary of how the topic has developed until now. Since these fluctuations are governed by the Goldstone director fluctuations (see e.g., \cite{PK77}, \cite{PP79}) they exist only in the nematic state. Applied external magnetic field suppresses the singular longitudinal fluctuations of the order parameter (similarly as it is the
case for the transverse director fluctuations, although with a different scaling over the magnetic field). Effectively
the reduction of the fluctuations changes the equilibrium value of the modulus of the order parameter in the
nematic state. Therefore it leads  to additional (with respect to the mean field contribution) fluctuational shift
of the nematic - isotropic ($N\, -\, I$) transition temperature. In subsection \ref{res} we calculate this shift and its dependence on the magnetic field $H$. Finally in subsection \ref{con} we summarize our results and confront them
with the available experimental data. Our prediction $\Delta T_c \propto H$ is in a qualitative
agreement with mentioned in \cite{ST16} almost linear scaling. Within our pure phenomenological
approach we are not in the position to estimate numeric prefactor in this
linear law, but perfection may be too much to expect from such a complicated system, where not easy to identify experimentally the various sources of the order parameter response to the external magnetic field.  The topic is still full of challenging open questions, and it may be fair to state, that the physical properties of bent-core liquid crystalline materials are far from being well described and understood.

 \subsection{Basic relations}
\label{bas}

As it is well known from any textbook on statistical physics or thermodynamics, (e.g., \cite{LL80}) 
there are two conditions to find the critical temperature for
the first order phase transition. For the particular case $I$ - $N$ transition, the conditions are
\begin{itemize}
\item
The both phases (with the equilibrium moduli of the orientational order parameter $S_1$ and $S_2$) have to be
minima of the free energy density (\ref{ntb2})
\begin{equation}
    \label{ntb4}
    \frac{\partial F}{\partial S_{1 , 2}} = 0\, ; \, \frac{\partial ^2 F}{\partial S_{1 , 2}^2} > 0
		    \, .
  \end{equation}
  
\item
At the transition temperature the free energies of the both phases are equal
\begin{equation}
    \label{ntb5}
    F(S_1) = F(S_2)
		    \, .
  \end{equation}
 \end{itemize}
 Without external magnetic field ($H=0$), these conditions (\ref{ntb4}), (\ref{ntb5}) applied to the
 free energy density (\ref{ntb2}) give two minima 
 \begin{equation}
 \label{ntb6}
   S_1 = 0\, ;\, S_2 = \frac{1}{6}\frac{\mu}{\lambda_4} + {\sqrt {\frac{1}{36}\frac{\mu^2}{\lambda_4^2} - \frac{2}{3}\frac{a}{\lambda_4}}} 
   \, ,
  \end{equation}
 and at the transition point
 \begin{equation}
 \label{ntb6}
   T_c = T_* + \frac{1}{27}\frac{\mu^2}{\alpha \lambda_4} 
   \, ,
  \end{equation}
  the equilibrium value of the modulus of the nematic orientational order is
  \begin{equation}
 \label{ntb7}
   S_{2 c}  = \frac{2}{9}\frac{\mu}{\lambda_4}
      \, .
  \end{equation}
 External magnetic field changes the equilibrium values of the moduli
 \begin{equation}
 \label{ntb8}
   S_{1 c} \to \frac{1}{3}\frac{\Delta \chi H^2}{a} \, ;\, S_{2 c} \to \frac{2}{9}\frac{\mu}{\lambda_4} + 9\frac{\lambda_4}{\mu^2}\Delta \chi H^2 - 6 \frac{\Delta a}{\mu} 
   \, ,
  \end{equation}
 where $\Delta a$ is the quantity we are interested in which determines $\Delta T_c(H)$.
 Combining everything we find in the main over $H^2$ approximation the known \cite{HE70} expression for the 
 mean-field shift of the transition temperature presented above (\ref{ntb3}) in terms of the Landau expansion (\ref{ntb2}).
 
 Thermal fluctuations add new ingredients and features to this mean-field picture (needed to be included since experimental data suggest noticeable deviations from mean-field behavior (\ref{ntb3})). First of all it is tempting to take into consideration critical fluctuations of the orientational order parameter $Q_{ik}$. For a weak first order phase transition these fluctuations lead
 to a non mean-field scaling (power law) behavior of the correlation and response functions. However, long wave-length
 (IR divergent) critical fluctuations do not produce the shift of the transition temperature. The wisdom borrowed from the standard phase transition theory \cite{LL80}, \cite{ST87} says that the transition temperature  $T_c$ has to be considered as a given phenomenological parameter (it is determined by the short wave-length, UV divergent contributions).
 To find the fluctuational field-dependent shift of the transition temperature $\Delta T_c(H)$, one has to find how magnetic field affects the module of the order parameter. Moreover if the fluctuation renormalization of the order parameter moduli are equal for the both states coexisting for the first order phase transitions at $T_c$ ($I$ and $N$
 in our case), in the main over $H$ approximation there is no shift of the transition temperature.  It is the case for the critical order parameter fluctuations, since in the critical region there is no difference between $I$ and $N$
 states.
 
 However in the nematic liquid crystals (as for any system with a broken continuous symmetry) in the entire
 range of the nematic state stability (not only in the critical region) there are soft (Goldstone) fluctuations of the director. The director fluctuations themselves do not shift of the transition temperature, but by virtue of the
 principle of the conservation of the modulus \cite{PP79}, the director fluctuations renormalize the longitudinal
 magnetic susceptibility (see e.g., \cite{FB73} for magnetic systems, or \cite{PK77} for nematic liquid crystals)  and thus effectively modify the module of the order parameter. The modification takes place
 only in the nematic phase, therefore the singular longitudinal fluctuations indeed lead to the field-dependent shift of the phase transition temperature.
 
\subsection{Longitudinal order parameter fluctuations}
\label{res}

 The longitudinal fluctuations of the orientational order parameter $Q_{ik}$ are fluctuations of the quantity
 $S$, the modulus of the order parameter. According to the fluctuation-dissipation theorem \cite{LL80},
 the longitudinal fluctuation correlation function is related to the longitudinal susceptibility (response 
 to an external magnetic field). The general theory of degenerate systems \cite{PP79} 
 states the principal of the conservation of the modulus, which if applied to the nematic liquid crystals 
 reads as
 \begin{equation}
 \label{ntb9}
   2 S \delta S = - \langle \delta {Q_{ik}^\perp }^2\rangle  
   \, ,
  \end{equation}
 where $\delta Q_{ik}^\perp $ is transverse (director) fluctuations of the order parameter.
 We conclude from (\ref{ntb9}) that strong transverse fluctuations give rise to weaker (but also singular in the long wave-length limit) longitudinal fluctuations. 
 From (\ref{ntb9}) one can find the longitudinal pair correlation function
 \begin{equation}
 \label{ntb10}
   \langle \delta S({\bf r})\, \delta S({\bf r}^\prime )\rangle  = \frac{1}{2 S^2} \langle \delta Q_{ik}^\perp ({\bf r})\,    \delta Q_{ik}^\perp ({\bf r}^\prime )\rangle ^2
   \, 
  \end{equation}
 or in the reciprocal (Fourier) space
 \begin{equation}
 \label{ntb11}
   \langle |\delta S({\bf q})|^2 )\rangle  = \frac{1}{2 S^2} \int \frac{d^3 p}{(2\pi )^3} \langle
   \delta Q_{ik}^\perp ({\bf p})\,    \delta Q_{mn}^\perp (-{\bf p})\rangle  \langle \delta Q_{ik}^\perp ({\bf p} + {\bf q})\,    \delta Q_{mn}^\perp (-{\bf p} - {\bf q})\rangle 
   \, . 
     \end{equation}
  For the standard Frank elastic orientational energy (which contains 3 elastic moduli for the uniaxial nematics
  and 15 moduli for the biaxial case (see e.g., \cite{BP81}	
	in an external magnetic field,
  the integral in (\ref{ntb11}) can be computed numerically. We present only the result in the one-constant $K$ approximation
  \begin{equation}
 \label{ntb12}
   \langle |\delta S({\bf q})|^2 )\rangle  \simeq \frac{T^2}{4 S^2 K^{3/2}(q^2 + \chi _a H^2)} \, , 
     \end{equation}
  where $\chi _a$ is the anisotropic part of the magnetic susceptibility.
  For a macroscopically large uniformly ordered sample (\ref{ntb12}) can be interpreted
  as mean squared longitudinal fluctuation
  \begin{equation}
 \label{ntb13}
   {\sqrt{\langle |\delta S|^2 \rangle }}   \equiv \delta S_{fl} = \frac{1}{V^{1/2}}\frac{T}{K^{3/4}} \frac{1}{S\chi _a^{1/4}}\frac{1}{\sqrt H} \,  
     \end{equation}
  Then the corresponding free energy density (i.e., per unit volume $V$) is
   \begin{equation}
 \label{ntb14}
  \delta F_{fl} \propto  \epsilon\frac{H}{S} \,  ,
     \end{equation}
  where for the sake of compactness we introduce notation
   \begin{equation}
 \label{ntb15}
   \epsilon = \frac{1}{2}\frac{T}{K^{3/4}} \frac{\Delta \chi }{\chi _a^{1/4}} \,  
     \end{equation}
  Performing again all calculations from the previous subsection \ref{bas} but with the longitudinal contribution
  (\ref{ntb14}) included, we end up (in the main approximation over $H$) with the following expression for the transition temperature shift
   \begin{equation}
 \label{ntb16}
   \Delta T_c(H) = \frac{243}{2\alpha }\frac{\lambda_4^3}{\mu^3}\epsilon H \,  .
     \end{equation} 
  This Exp. (\ref{ntb16}) is our main result in the work, which is ready to be confronted with experimental data.
  Unfortunately scanning the literature we did not find systematic measurements for $\Delta T_c(H)$ for the
  bent-core mesogen liquid crystals. Therefore for the time of writing of the manuscript Exp. (\ref{ntb16})
  has to be considered as our theoretical predictions to be checked experimentally.

To conclude this section we repeat our main findings. We study a macroscopic counterpart to the microscopic mechanism of the straightening dimer mesogens conformations \cite{ST16}. The proposed mechanism enables to explain qualitatively recent experimental observations of the unprecedentedly large shift of the nematic - isotropic transition temperature. 
Our interpretation is based on singular longitudinal fluctuations of the nematic order parameter. Since these fluctuations are governed by the Goldstone director fluctuations they exist only in the nematic state. Applied external magnetic field suppresses the singular longitudinal fluctuations of the order parameter (similarly as it is the
case for the transverse director fluctuations, although with a different scaling over the magnetic field). Effectively
the reduction of the fluctuations changes the equilibrium value of the modulus of the order parameter in the
nematic state. Therefore it leads  to additional (with respect to the mean field contribution) fluctuational shift
of the nematic - isotropic transition temperature. Our mechanism works for any nematic liquid crystals, however the magnitude of the fluctuational shift increases with decrease of the Frank elastic moduli. Since some of these moduli supposed to be anomalously small for so-called bent-core or dimer nematic liquid crystals (see e.g., \cite{DO01}, \cite{MD16}), just these kinds of the mesogenes are promising candidates for the observation of the predicted fluctuation shift of the phase transition temperature.

There are some issues  we have not tackled here.
For example, the physical mechanism we are advocating in this work is designed to emphasize that singular longitudinal
fluctuations lead to the external field induced shift of the transition temperature. We ignore all other types of the fluctuations that may ultimately produce a more accurate description of the $\Delta T_c(H)$ behavior.  A more specific study of all these phenomena might become appropriate should suitable experimental results become available.
In our opinion one important lesson to be learned from our work is that independent of the precise values of the model parameters (which, especially for new liquid crystalline materials are yet unknown experimentally) the general tendencies and forms of the fluctuational contributions into $\Delta T_c(H)$  can be predicted theoretically and tested experimentally.

\section{Outlook, Conclusions and Perspectives.}
\label{con}

The presented above material does not provide a comprehensive review devoted to the $N_{TB}$
nematic liquid crystals. It is rather a personal view of the author, although informed
on the known fro the literature achievements in the field, but mainly based on his own
contribution. In the absence of elementary and pedagogical textbooks or reviews,
it seems appropriate at this time to take a fresh look at the ream of $N_{TB}$ liquid crystals.
It is hoped that this review provides a useful and self-contained starting point for new
researchers entering the field.

With this aim in mind, we review structural and physical properties of only one new (relatively recently discovered) type of liquid crystals, so-called twist-bend nematics $N_{TB}$.
Correspondingly to the topic we describe the general features of the $N_{TB}$ nematics,
and present Landau Theory of the easy axis $N_{TB}$ nematics. 
This Free energy (supplemented by the Frank elastic energy), looks similar to the Free energy expansion for the nematic - smectic $A$ phase transition. However in our model 
the order parameter $\bm \psi $ has $n = 4$ components  in three-dimensional
space. In such a case a stable fixed point can be accessible only for
very large number of components of the order parameter ($n > 238$). 
Therefore it is expected a first-order transition. 
This prediction is in agreement with experimental data.
If this transition is a weak 1-st order phase transition,
fluctuation effects can be observed only in a narrow vicinity of the phase transition. 
 
Special sections are 
devoted to a description of light scattering in the $N_{TB}$ liquid crystals.
In the vicinity of the $N\, -\, N_{TB}$ phase transition, in the $N$ state, we 
include into consideration five soft degrees of freedom. 
Namely, two Goldstone-like components of the director, and three
short-range correlated components of the
vector order parameter ${\bm \varphi }$. These five degrees of freedom corresponds to the five modes (two hydrodynamic and three pseudo-hydrodynamic ones) which can be detectable in the light scattering experiments. Similarly from the $N_{TB}$ side
there are six soft degrees of freedom, forming six modes potentially observable in the light scattering experiments.
Namely, one hydrodynamic mode, describing phase of the long ranged order parameter fluctuations, and five 
quasi-hydrodynamic modes. The matter is that in the $N_{TB}$ state the director orientation is coupled 
to the vector order parameter ${\bm \varphi}$. Thus the director and ${\bm \varphi}$ orientation
fluctuations acquire an energy gap $\omega \neq 0$ at ${\bf q} \to 0$.
Nevertheless in the vicinity of the $N$ -  $N_{TB}$ phase transition these modes are not too fast (the gap in
the excitation spectra is relatively small), therefore the modes can be detected in the light scattering experiments.
It is worth to noting that because the optical wave vector is smaller than the inverse 
pitch of the $N_{TB}$ heliconical structure, only second order scattering 
processes (proportional to the square of the $N_{TB}$ order parameter fluctuations) contribute to the light scattering intensity. 

Rheological behavior of the ordered $N_{TB}$ samples is also a topic considered in this
work.
We show that coarse grained dynamic description of the $N_{TB}$ phase
allows to rationalized all observed experimentally rheological regimes. 
Based on such coarse grained description (supplemented by the maximum rate of entropy production principle) we estimated
the critical values of the shear rate separating different (shear-thinning, and quasi-Newtonian)rheological regimes. We presented scaling analysis of  three regimes of a steady shear  viscosity
curve which is in qualitative agreement with previously reported observations.
We hope that presented results stimulate
discussions on the intriguing and important issues of non-Newtonian rheology in liquid crystals.

Finally we analyze also how external d.c. electric field shifts $N$ - $N_{TB}$ phase transition point.
We claim that singular longitudinal fluctuations of the nematic order parameter is a macroscopic counterpart to the microscopic mechanism of the straightening dimer mesogens conformations, observed
experimentally. The proposed mechanism enables to explain experimental observations of the unprecedentedly large shift of the nematic - isotropic transition temperature.

There are many issues  we have not touched in this work, and as well there are some physical ingredients missed in our approach. 
To name a few it is worth to mention typical for any soft matter systems, non-linear, non-local, far-from-equlibrium and interfacially
elastic phenomena. We plan to address these issues in a future work extending the phenomenological methodology implemented here.
We hope that although our presentation in this work is mainly heuristic and representative (rather than encyclopedic),
this simple introduction could inspire readers to undertake a more scholarly investigation of $N_{TB}$ liquid crystals physics.


\begin{thebibliography}{99}


\bibitem{SD04}
T.J.Sluckin, D.A.Dunmur, H.Stegemeyer, Eds., Crystals That Flow: Classic Papers from the History of Liquid Crystals,
Liquid Crystals Book Series, Taylor and francis Group, London (2004).


\bibitem{GP93} P. G. de Gennes, J. Prost,
 The physics of liquid crystals,
 Clarendon Press, Oxford (1993).

\bibitem{KL06} M.Kleman, O.Lavrentovich Soft Matter Physics: An Introduction, Springer, Berlin (2003).


\bibitem{OP06} P.Oswald, P.Pieranski, Smectics and columnar liquid crystals, Taylor and Francis, New York (2006).


\bibitem{KH07}
I-C. Khoo, Liquid Crystals (2d edition), Wiley, New York (2007).

\bibitem{BL11}
L.M. Blinov, Structure and properties of liquid crystals, Springer, New York (2011).


\bibitem{DZ12}
P.V. Dolganov, V.M. Zhilin, V.K. Dolganov, E.I. Kats, Phys. Rev. E {\bf 86}, 020701(R) (2012).

\bibitem{DZ12a}
P.V. Dolganov, V.M. Zhilin, E.I. Kats, JETP {\bf 113}, 1140 (2012).

\bibitem{DK13}
P.V. Dolganov, E.I. Kats, Liquid Crystals Reviews, {\bf 1}, 127 (2013). 

\bibitem{HW15} 
C.C.Huang, S.Wang, L.Pan, Z.Q.Liu, B.K.McCoy, Y.Sasaki, K.Ema, P.Barois, R.Pindak, Liq. Cryst. Rev. {\bf 3}, 58
(2015).




\bibitem{FV21}
A.Fukuda, J.K.Vij, Y.Takanishi, Phys. Rev. E, {\bf 104}, 014705 (2021).


\bibitem{HS09}
 L.E. Hough, M. Spannuth, M.Nakata, et al., Science, {\bf 325}, 452 (2009).

 \bibitem{PN10}
 V.P.Panov, M.Nagaraj, J.K.Vij, et al., Phys. Rev. Lett., {\bf 105}, 167801 (2010).

 \bibitem{CD11}
 M.Cestari, S.Diez-Berart, D.A.Dunmur, et al., Phys. Rev. E, {\bf 84}, 031704 (2011).

 \bibitem{BK13}
 V.Borshch, Y.K.Kim, J.Xiang, et al., Nature Commun., {\bf 4}, 2635 (2013).

 \bibitem{MD14}
 J.Mandle, E.J.Davis, S.A.Lobato, et al., Phys. Chem. Chem. Phys., {\bf 16}, 6907 (2014).


\bibitem{KL14} 
E.I. Kats, V.V. Lebedev, JETP Lett., {\bf 100}, 110 (2014).

\bibitem{KA17}
E.I.Kats, Low Temperature Physics (Fizika Nizkikh Temperatur), {\bf 43}, 7 (2017).



\bibitem{SW76}
J.Swift, Phys. Rev. A {\bf 14}, 2274 (1976).


 

\bibitem{BR75} S.A.Brazovskii, JETP, {\bf 41}, 85 (1975) [ZhETF,  {\bf 68}, 175 (1975)].

\bibitem{KLM93}  
E.I. Kats, V.V. Lebedev, A.R. Muratov,
Physics Reports {\bf 228}, 1 (1993).


\bibitem{WK74}
K.G.Wilson, J.Kogut, Phys. Reps., {\bf 12}, 75 (1974).





\bibitem{PP79}
A.Z. Patashinskii, V.L. Pokrovskii,
Fluctuation Theory of Phase Transitions, Pergamon Press, New York (1979).



 \bibitem{HL74}
 B.I.Halperin, T.C.Lubensky, S-K.Ma, Phys. Rev. Lett., {\bf 32}, 292 (1974).

 \bibitem{LC78}
 T.C.Lubensky, J-H. Chen, Phys. Rev. B, {\bf 17}, 366 (1978).


\bibitem{LL84}
L.D.Landau, E.M.Lifshitz, Electrodynamics of Continuous Media,
Volume 8 in Course of Theoretical Physics, Second Edition, Pergamon Press, New York (1984).



\bibitem{KA12}
 S.Kaur, J.Addis, C.Greco, et al., Phys.Rev. E, {\bf 86}, 041703 (2012).


 \bibitem{ML13}
 C.Meyer, G.R.Lukhurst, I.Dozov, Phys. Rev. Lett., {\bf 111}, 067801 (2013).


\bibitem{PS16}
Z.Parsouzi, S.M.Shamid, V.Borshch, P.K.Challa, A.R. Baldwin, M.G. Tamba, C.Welch, G.H.Mehl,
J.T.Gleeson, A.Jakli, O.D.Lavrentovich, D.W.Allender, J.V. Selinger, S.Sprunt,
Phys. Rev., X {\bf 6}, 021041 (2016).

\bibitem{KK20} M.P.Kumar, P.Kula, S.Dhara, 
Phys. Rev. Materials {\bf 4}, 115601 (2020).

\bibitem{KK21}
M.Praveen Kumar, J.Karcz, P.Kula, S.Dhara, Phys. Rev. Materials, {\bf 5}, 115605 (2021).


\bibitem{KA22}
E.I.Kats, 
JETP Letters, {\bf 116}, 255-260 (2022).


\bibitem{KL93}
 E.I.Kats and V.V.Lebedev,
 Fluctuational Effects in the Dynamics of Liquid Crystals, Springer, 1993.



\bibitem{SK14} 
S.M.Salili, C.Kim, S.Sprunt, J.T.Gleeson, O.Parric, A.Jakli, 
RSC Adv., {\bf 4}, 57419 (2014). 





\bibitem{MD16}
C.Meyer, I.Dozov,
Soft Matter, {\bf 12}, 574 (2016).




\bibitem{KL86}
E.I.Kats, V.V.Lebedev, JETP, {\bf 64}, 518 (1986) [ZhETF, {\bf 91}, 871 (1986)].

\bibitem{MP70}
P.C.Martin, P.S.Pershan, J.Swift, Phys.
Rev. Letters {\bf 25}, 844 (1970).

\bibitem{MP72}
P.C.Martin, O.Parodi, P.S.Pershan, Phys. Rev. A, {\bf 6}, 2401 (1972).


\bibitem{SB10}
D. Svensek, H.R. Brand, Adv. Polym. Science, Layered Systems Under Shear Flow, 
Adv. Polym. Science, Springer-Verlag, Berlin Heidelberg (2010).


\bibitem{FK14}
S.Fujii, S.Komura, Ch.-Yi D.Lu, 
Materials, {\bf 7}, 5146 (2014).



\bibitem{PA95}
P.Panizza, P.Archabault, D.Roux, J. de Phys. II, France, {\bf 5}, 303 (1995).



\bibitem{LL80}
 L.D.Landau, E.M.Lifshitz, Course of Theoretical Physics, Statistical Physics, Part 1, Pergamon Press, New York (1980).





 \bibitem{ST87}
 H.E.Stanley, Introduction to phase transitions and critical phenomena,
 Oxford University Press, New York, 1987.


\bibitem{HU87}
K.Huang, Statistical Mechanics, 2nd edition, John Wiley and Sons, Montreal (1987).


\bibitem{CL00}
P. M. Chaikin, T. C. Lubensky, 
Principles of Condensed Matter Physics,                                          
Cambridge, Cambridge University Press (2000).


\bibitem{PI91}
S.A.Pikin, Structural Transformations in Liquid Crystals, Gordon and Breach, New York (1991).

\bibitem{AN91}
 M.A.Anisimov, Critical phenomena in liquids and liquid crystals,
 Gordon and Breach, Philadelphia, 1991.


\bibitem{CH92}
S. Chandrasekhar, Liquid Crystals,
Cambridge, Cambridge University Press (1992).



\bibitem{BC94} L.M.Blinov, V.G.Chigrinov, Electrooptic Effects in Liquid Crystals, Springer,
New York (1994).

\bibitem{OP05}
P.Oswald, P.Pieranski, Nematic and Cholesteric Liquid Crystals: Concepts and Physical Properties Illustrated
by Experiments, Liquid Crystals Book Series, Taylor and Francis, London (2005).




\bibitem{HE70} W.Helfrich, Phys. Rev. Lett., {\bf 24}, 201 (1970).


\bibitem{ST16} S.M.Saliti, M.G.Tamba, S.N. Sprunt, C.Welch, G.H.Mehl, A.Jakli, J.T.Gleeson, Phys. Rev. Lett., {\bf 116}, 217801 (2016).

\bibitem{OW08} T.Ostapenko, D.B.Wiant, S.N.Sprunt, A.Jakli, J.T.Glisson, Phys. Rev. Lett., {\bf 101}, 247801 (2008).

\bibitem{FV11} O.Francescangeli, F.Vita, F.Fauth, E.T.Samulski, Phys. Rev. Lett., {\bf 107}, 207801 (2011).

\bibitem{TS13} T.B.T. To, T.J.Sluckin, G.R.Luckhurst, Phys. Rev. E, {\bf 88}, 062506 (2013).

\bibitem{FB73} M.E.Fisher, N.B.Barber, D.Jasnow, Phys. Rev. A, {\bf 8}, 1111 (1973).

\bibitem{PK77} V.L.Pokrovskii, E.I.Kats, Sov. Phys. JETP, {\bf 46}, 405 (1977) (ZhETF, {\bf 73}, 774 (1977)).


\bibitem{BP81}     H.Brand, H.Pleiner, Phys. Rev. A, {\bf 24}, 2777 (1981).


 \bibitem{DO01}
 I.Dozov, Europhys. Letters, {\bf 56}, 247 (2001).




\end{thebibliography}
\end{document}